# Noise reduction in plasmonic amplifiers

Andrey A. Vyshnevyy and Dmitry Yu. Fedyanin*

*Laboratory of Nanooptics and Plasmonics, Moscow Institute of Physics and Technology, Dolgoprudny 141700, Russian Federation*

E-mail: dmitry.fedyanin@phystech.edu

Surface plasmon polariton amplification gives the possibility to overcome strong absorption in the metal and design truly nanoscale devices for on-chip photonic circuits. However, the process of stimulated emission in the gain medium is inevitably accompanied by spontaneous emission, which greatly increases the noise power. Here we present an efficient strategy for noise reduction in plasmonic amplifiers, which is based on gain redistribution along the amplifier. We show that even a very little gain redistribution (~3%) gives the possibility to increase the signal-to-noise ratio by about 100% and improve the bit error ratio by orders of magnitude.

In the past decade, advances in nanophotonics and plasmonics allowed to bring optical communications to the nanoscale where they can meet the exponentially growing demand for the communication speed and throughput.[1–4] The ability of signal amplification is one of the key requirements for reliable data transmission and detection.[5] At the nanoscale, optical amplification can be provided by active nanophotonic[6,7] or plasmonic structures.[8–11] The latter give the opportunity to bridge the scale gap between on-chip electronic and optical components. However, this opportunity comes at a price of increased internal optical losses attributed to the absorption in the metal, which is an essential part of plasmonic devices. The modal loss rapidly increases as the mode confinement gets stronger.[12,13] Nevertheless, a remarkable progress on plasmonic nanolasers and amplifiers[9,10,14–19] shows that the net optical amplification can be achieved even in plasmonic waveguides with deep-subwavelength confinement. However, the problem is that every optical amplifier adds spontaneously emitted photons to the signal-carrying mode.[20] The stochastic nature of the spontaneous emission process manifests itself in a photonic noise producing photocurrent fluctuations at the photodetector.[21,22] Since plasmonic amplifiers operate at very high degrees of population inversion,[19] the spontaneous emission rate is much higher than in photonic amplifiers, so is the noise power.[23,24] This strong noise leads to errors at the receiver side, which can greatly decrease the information capacity of the data transmission channel and render it useless.

Here, we present a strategy to greatly improve the noise characteristics of nanoscale plasmonic amplifiers by controlling the spatial distribution of the modal gain along the amplifier. The proposed method of noise reduction does not affect the net amplifier gain and can be easily implemented in practical devices, which we numerically demonstrate on the example of a practical metal/semiconductor amplifier.

Noise added by an optical amplifier is produced by intensity fluctuations due to the interference between spectral components of the signal propagating in the amplifier and the amplified spontaneous emission (ASE) generated by the gain medium. Depending on what spectral components interfere, two contributions can be distinguished: the signal-spontaneous beat noise and the spontaneous-spontaneous beat noise.[20,21] In contrast to optical fiber amplifiers, nanoscale active plasmonic structures feature a very high ASE power, which can significantly exceed the power of the transmitted signal. Therefore, the power of the spontaneous-spontaneous beat noise, which does not depend on the signal power, can be much stronger than that of the signal-spontaneous beat noise.[23] A narrow-band pass filter between the amplifier and receiver can solve this problem reducing the noise power density $S$ to that determined only by the interference of the signal and the ASE at the carrier frequency of the signal $v_0$:[23]

$$S(f) = \left(\frac{4e^2}{hv_0^2} P_{\text{out}}\right) p_{\text{sp}}(hv_0), \quad (1)$$

where, $h$ is the Planck constant, $e$ is the elementary charge, $p_{\text{sp}}(hv_0)$ is the spectral power density of amplified spontaneous emission at the output of the amplifier, $P_{\text{out}}$ is the signal power at the output of the amplifier and $f$ is the noise frequency, which is limited by the RC rise time of the receiver circuit (~0.1 ns). $p_{\text{sp}}(hv_0)$ greatly depends on the distribution of the modal loss $\alpha(z)$ and modal gain $G(z)$ along the amplifier. The rate of spontaneous emission of the gain medium into the signal-carrying mode per unit of the amplifier length is equal to[23]

$$R_{\text{sp}}(hv_0, z) = \frac{2}{h} G(z) N_{\text{sp}}(z), \quad (2)$$

where $N_{sp}$ is the spontaneous emission factor of the gain medium. $N_{sp}$ is defined as a ratio of the spontaneous emission rate into the signal-carrying mode to the stimulated emission rate per one photon in the same mode. In subwavelength plasmonic waveguide structures, the propagation loss is typically as high as 500 – 5000 cm$^{-1}$ (Refs. 9,10,25). Therefore, one has to provide a very high gain $G > \alpha$, which can be created only under strong population inversion. However, a positive material gain in the active medium is unavoidably accompanied by strong spontaneous emission. The higher the population inversion, the higher the spontaneous emission rate. For example, in a semiconductor active medium, $N_{sp} = 1/\left[1 - \exp\left(\frac{h\nu_0 - (F_e - F_h)}{k_B T}\right)\right]$, where the difference $(F_e - F_h)$ between quasi-Fermi levels characterizes the population inversion. At a high material gain, $(F_e - F_h - h\nu_0) \gtrsim 2k_B T$ and, therefore, $N_{sp} \approx 1$. The spectral power density of the amplified spontaneous emission at the output of the amplifier can be found by integrating $R_{sp}(h\nu_0)$ over the amplifier length. Here, we should note that the spontaneously emitted surface plasmons are also amplified by the plasmonic amplifier, which adds the propagation factor[23] $\exp\left(\int_z^L [G(z') - \alpha(z')]dz'\right)$, and only half of the spontaneously emitted surface plasmons propagate towards the output. Finally, taking into account that each surface plasmon quantum carries the energy $h\nu_0$, we obtain the expression for the ASE spectral power density at $z = L$:

$$p_{sp}(h\nu_0) = \int_0^L \nu_0 G(z) N_{sp}(z) \exp\left(\int_z^L [G(z') - \alpha(z')]dz'\right) dz. \qquad (3)$$

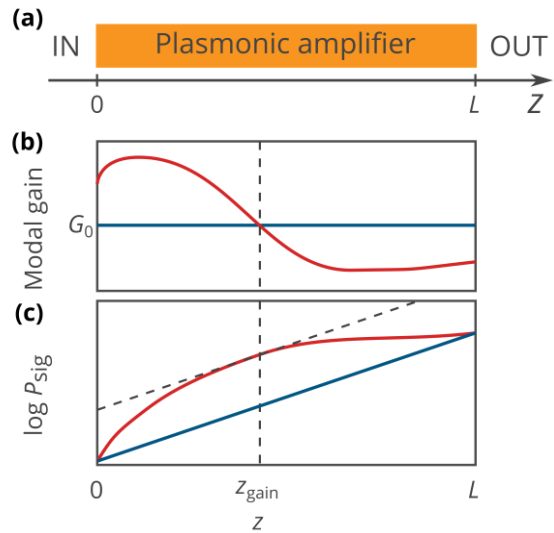

**Fig. 1.** (a) Schematic illustration of a plasmonic amplifier. (b) Two possible distributions of the modal gain in the amplifier at a fixed net amplifier gain $P_{out}/P_{in}$: uniform (blue curve) and non-uniform (red curve). (c) Qualitative log-scale plot of the signal power $P_{sig}(z)$ inside the plasmonic amplifier for two spatial gain distributions shown in panel b. Dashed lines serve as a guide for the eye.

The modal gain $G(z)$ provided by the active medium is determined by the pump rate. Therefore, one can control $G(z)$, which is especially easy to do in practical devices, which should be pumped electrically.[26,27] This possibility gives one extra degree of freedom, which can be used to enhance the amplifier performance, particularly to minimize the noise. Can we improve the noise characteristics of the amplifier by slightly redistributing the modal gain $G(z)$ along the amplifier so that the net amplifier gain $P_{out}/P_{in} = \exp\left(\int_0^L [G(z) - \alpha(z)]dz\right)$ remains the same?

Let us consider a single mode amplifier with a redistributed modal gain $G(z)$ (see Fig. 1). We also assume the modal loss $\alpha(z)$ to be uniformly distributed along the amplifier, which is almost always valid for nanoscale plasmonic waveguides, where losses are determined by absorption in the metal. By replacing the integral $\int_z^L [G(z') - \alpha]dz'$ in Eq. (3) by $P_{out}/P_{sig}(z)$ where $P_{sig}(z)$ appears to be the power of the signal at a distance $z$ from the input of the amplifier, we obtain

$$p_{sp}(h\nu_0) = \nu_0 \left(\frac{P_{out}}{P_{in}} - 1\right) + \nu_0 \int_0^L \frac{\alpha P_{out}}{P_{sig}(z)} dz. \qquad (4)$$

The second term in Eq. (4) is a function of $G(z)$, since $P_{sig}(z) = P_{in} \exp\left(\int_0^z [G(z') - \alpha]dz'\right)$. If the gain is uniformly distributed along the amplifier $G(z) = G_0$,

$$p_{sp}(h\nu_0) = \nu_0 \left(\frac{P_{out}}{P_{in}} - 1\right) \frac{G_0}{G_0 - \alpha}. \qquad (5)$$

Due to the high propagation loss of strongly confined plasmonic modes, the net modal gain $G_0 - \alpha$ is much lower than $G_0$. Therefore, the noise power density in plasmonic amplifiers is much higher than in photonic amplifiers at the same net amplifier gain $P_{out}/P_{in}$. However, since $(G(z) - \alpha) \ll G(z)$, even a small variation of $G(z)$ can give the possibility to significantly decrease the noise. To achieve this, one should redistribute the modal gain in such a way that $P_{sig}(z)$ is higher than $P_{in} e^{(G_0 - \alpha)z}$ for every $z$. This condition can be satisfied if $G(z) > G_0$ at $z < z_{gain}$ and $G(z) < G_0$ at $z > z_{gain}$ (see Fig. 1). We stress again that the net amplifier gain $P_{out}/P_{in}$ is unaffected.

If the net amplifier gain is fixed, then the shorter the amplification section, the lower the noise at the output. Ideally, the gain distribution should approach $G(z) = G_0 L \delta(z-l)$, where $l \ll L$, i.e., the signal should be amplified in a very short region near the input of the amplifier $0 \leq z \leq 2l$ and then propagate in a passive section towards the amplifier output. This gives us the fundamental limitation of the signal-spontaneous beat noise:

$$S(f) = \left(\frac{4e^2}{h\nu_0} P_{\text{out}}\right) \left[\frac{P_{\text{out}}}{P_{\text{in}}} - \exp(-\alpha L)\right] \quad (6)$$

The problem is that in nanoscale plasmonic structures, the modal loss $\alpha$ is very high. Therefore, it is not possible to create a modal gain much higher than $G_0 = \alpha$. Nevertheless, combining a stronger amplification in the first half of the amplifier ($G_0 + \Delta G$) and a weaker amplification ($G_0 - \Delta G$) in the second half (Fig. 2), it is possible to reduce the noise power.

Let us consider a realistic truly nanoscale plasmonic amplifier and evaluate the influence of the gain redistribution on its noise characteristics. Figure 2(a) shows a cross-section of the amplifier based on the active T-shaped plasmonic waveguide.[27,28] The substrate is made of a ternary $AlAs_ySb_{1-y}$ alloy lattice matched to the $In_xGa_{1-x}As$ gain medium. The $In_xGa_{1-x}As/AlAs_ySb_{1-y}$ heterojunction is used to confine injected electrons and holes to the active region. $x = 0.52$ is chosen to match the peak wavelength of the gain spectrum of $In_xGa_{1-x}As$ to the free space signal wavelength $\lambda = 1550$ nm. A 2-nm-thick insulator layer (such as $HfO_2$)[28] between the metal and semiconductor is required to efficiently inject electrons from the metal through a tunnel metal-insulator-semiconductor junction.[28,29] Copper is chosen to form a surface plasmon supporting interface due to its outstanding optical properties at $\lambda = 1550$ nm (Refs. 30,31). The waveguide shown in Fig. 2(a) is single-mode and supports only the fundamental plasmonic mode strongly confined to the $Cu/HfO_2/InGaAs$ contact,[28] which is confirmed by direct finite element simulations using COMSOL Multiphysics (Comsol Inc., Burlington, MA). The modal loss $\alpha$ is found to be 970 cm$^{-1}$ and the mode confinement factor to the active region[32] equals $\Gamma = 1.09$. The moderate value of the modal loss among other semiconductor plasmonic waveguides with similar dimensions predetermines a good noise performance.[33] Such an active plasmonic waveguide can be pumped either optically or electrically. However, a detailed discussion of the pump mechanisms is out of the scope of the present letter and will be published elsewhere. Here, we focus on the noise characteristics. We start by considering the amplifier with $P_{\text{out}}/P_{\text{in}} = 1$, which corresponds to the case when high propagation losses ($\alpha = 970$ cm$^{-1}$) are fully compensated by the gain in the semiconductor ($G_0 = \alpha$). This regime enables high-speed data transmission over a long distance via a truly nanoscale waveguide.[27,34] The amplifier length is set to be $L = 1$ mm, which corresponds to a typical length of global on-chip interconnects.[2,35]

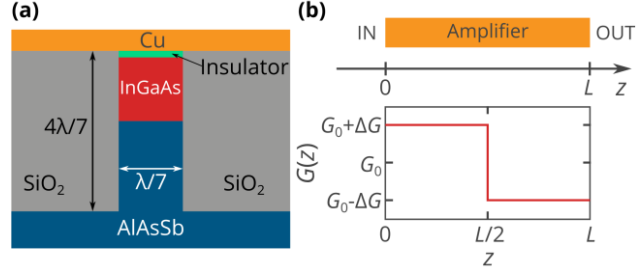

**Fig. 2.** (a) Cross-section of the plasmonic amplifier. (b) Spatial distribution of the modal gain along the plasmonic amplifier.

The ASE spectral power density, which determines the noise power, is directly calculated from the simulated material gain spectrum and spontaneous emission rate in InGaAs, which are found using the densities of non-equilibrium electrons and holes (for methods, see Ref. 23,28). Figure 3(a) clearly shows that the noise power decreases with $\Delta G$. At $\Delta G = 1000$ cm$^{-1}$ ($\Delta G/G_0 \approx 100\%$), the beat noise is reduced by a factor of 42 from the initial level and reaches a noise level that is only 4.7 times higher than the fundamental limit established by Eq. (6). On the other hand, if we exchange the amplifying and attenuating halves of the amplifier (i.e. change the sign of $\Delta G$ to negative) the noise power at the output significantly increases [Fig. 3(a)]. At a relatively small $\Delta G = -100$ cm$^{-1}$ ($\Delta G/G_0 \approx -10\%$), the noise is 30 times stronger than at $\Delta G = 0$. Notably, this noise increase is much stronger than the noise decrease at $\Delta G = 100$ cm$^{-1}$ ($\Delta G/G_0 \approx +10\%$) [Fig. 3(a)].

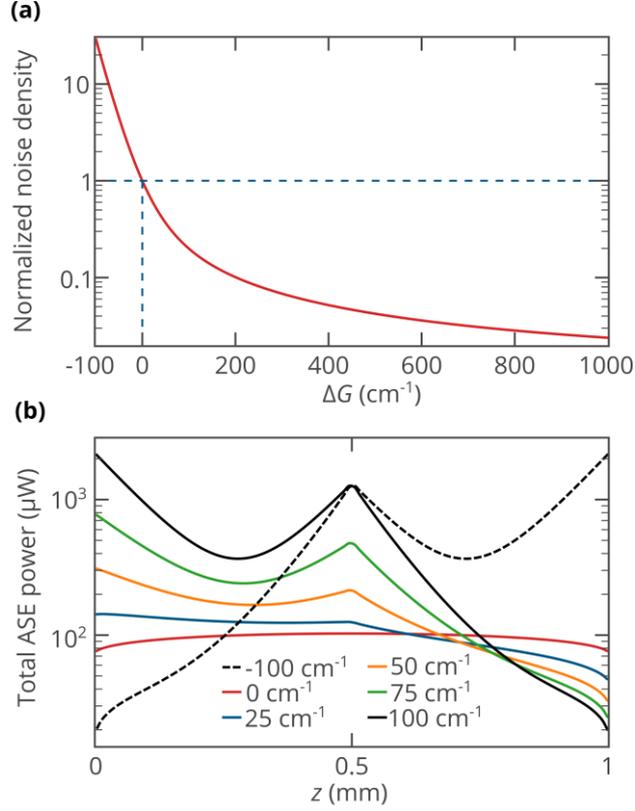

**Fig. 3.** (a) Noise density at the output of the plasmonic amplifier shown in Fig. 2, which is normalized to the noise of the amplifier with a uniform gain distribution ($G(z) = G_0 = 970$ cm$^{-1}$). (b) Spatial distribution of the total ASE power defined as the sum of the powers of the forward and backward propagating ASE inside the amplifier for different values $\Delta G$ (see Fig. 2(b) for the gain profile).

It is important to note that by redistributing the modal gain, one also changes the spontaneous emission rate (see Eq. (2)) and, consequently, the ASE power at the output of the amplifier. By integrating Eq. (3) over the whole spectra of spontaneous emission and modal gain and taking into account the fact that spontaneously emitted surface plasmon quanta propagate in both directions of the amplifier, we obtain a spatial distribution the ASE power in the amplifier [Fig. 3(b)]. The ASE power at the output of the amplifier decreases as $\Delta G$ increases, which is beneficial for practical applications [Fig. 3(b)]. However, the ASE power inside the amplifier increases with $\Delta G$. At $\Delta G = 100$ cm$^{-1}$ ($\Delta G/G_0 \approx 10\%$), it reaches 2 mW, which is 20 times higher than the maximum ASE power at $\Delta G = 0$ (100 µW). Such a high ASE power is extremely undesirable since the ASE depletes the carriers densities in the active medium[20]) and therefore increases the power consumption. Moreover, the ASE is absorbed in the metal, which increases the heat generation rate in the amplifier.[36])

Despite that it is difficult for the amplifier to operate at high $\Delta G$ due to the high ASE power in the amplifier, even a little gain redistribution is beneficial for the design of practical devices. $\Delta G$ as low as 25 cm$^{-1}$ ($\Delta G/G_0 \approx 2.6\%$) gives the possibility to reduce the noise power by 40%. In this case, the maximum ASE power is only 30% higher than in the amplifier with no gain redistribution. It is noteworthy that by placing an attenuating part ($G < G_0$) before the amplifying part ($G > G_0$), one only impairs the amplifier performance since both the ASE power and noise level are increased compared to the amplifier with a homogeneous gain distribution (Fig. 3).

The noise power is directly related to the bit error ratio (BER), which is the percentage of bits transmitted or detected incorrectly. Since $\log(\text{BER}) \propto 1/S$ (Refs. 21,33), a twofold reduction in the noise power corresponds to at least 4 orders of magnitude improvement of the BER in low-quality communication channels (BER ~ $10^{-5}$). On the other hand, if the quality of the communication channel is high at $\Delta G = 0$ (BER < $10^{-9}$), the BER decreases by 8 orders of magnitude when we slightly ($\Delta G/G_0 \approx 3\%$) redistribute the modal gain in the plasmonic amplifier.

The strategy for noise reduction based on the gain redistribution can be applied with no change to amplifiers with any net amplification coefficient $P_{out}/P_{in}$ (Fig. 4). However, a remark should be made. As the net amplifier gain increases, the influence of the gain redistribution on the noise power slightly decreases (Fig. 4).

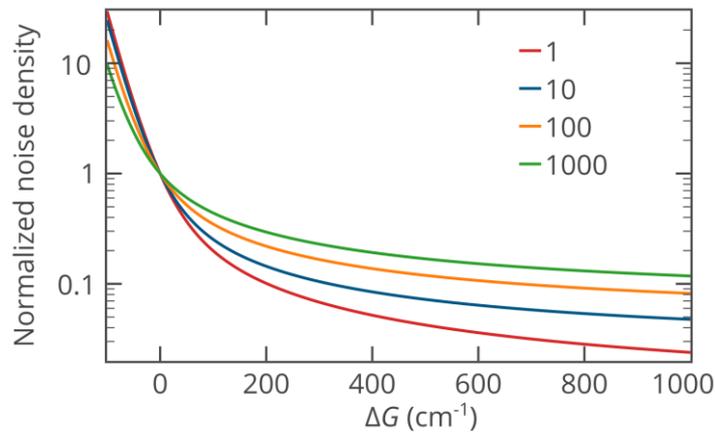

**Fig. 4.** Noise density at the output of the amplifier with gain redistribution, which is normalized with respect to the noise at the output of the amplifier with a uniform gain

distribution, for different values of the net amplifier gain $P_{out}/P_{in}$.

In summary, we have investigated for the first time plasmonic amplifiers with a modal gain non-uniformly distributed along the amplifier and evaluated the influence of the gain distribution on the noise characteristics. We have shown that the noise power can be greatly reduced by slightly increasing the modal gain in the first half of the amplifier and decreasing it in the second half of the amplifier at a fixed net amplification coefficient $P_{out}/P_{in}$. Despite that a strong gain redistribution required for noise suppression is difficult to be implemented in practical devices due to a very high ASE power in the amplifier, even a very little gain redistribution ($\Delta G/G_0 \approx 3\%$) gives the possibility to reduce the noise power by about 50%. Since the bit error ratio of the communication depends exponentially on the signal-to-noise ratio, such a noise reduction increases the reliability of communication by more than four orders of magnitude. These numbers can be further improved by optimizing the modal gain distribution. However, the optimization should be performed with respect to both the energy efficiency and the signal-to-noise ratio, since the increased ASE power greatly increases the power consumption of the amplifier. Thus, the proposed strategy for noise reduction allows to improve the noise characteristics of plasmonic amplifiers "at almost no cost", which opens new avenues in the design and development of nanoscale optical amplifiers for deep-subwavelength nanophotonic circuits.

**Acknowledgments**

Derivation of the noise characteristics (A.A. Vyshnevyy) is supported by the RSF (project no17-79-10488), numerical simulations of the semiconductor plasmonic amplifiers (D.Yu. Fedyanin) are supported by the RSF (project no17-79-20421).
**References**
1) D.A.B. Miller, Appl. Opt. **49**, F59 (2010).
2) C. Sun, M.T. Wade, Y. Lee, J.S. Orcutt, L. Alloatti, M.S. Georgas, A.S. Waterman, J.M. Shainline, R.R. Avizienis, S. Lin, B.R. Moss, R. Kumar, F. Pavanello, A.H. Atabaki, H.M. Cook, A.J. Ou, J.C. Leu, Y.-H. Chen, K. Asanović, R.J. Ram, M.A. Popović, and V.M. Stojanović, Nature **528**, 534 (2015).
3) D.A.B. Miller, Proc. IEEE **88**, 728 (2000).
4) B. Jalali and S. Fathpour, J. Lightwave Technol. **24**, 4600 (2006).
5) O. Wada, *Optoelectronic Integration: Physics, Technology and Applications* (Springer Science & Business Media, New York, 2013).

6) G. Roelkens, Frontiers in Optics 2016, FF5F.4.
7) H. Park, A.W. Fang, O. Cohen, R. Jones, M.J. Paniccia, and J.E. Bowers, IEEE Photonics Technol. Lett. **19**, 230 (2007).
8) J.S.T. Smalley, F. Vallini, Q. Gu, and Y. Fainman, Proc. IEEE **104**, 2323 (2016).
9) N. Liu, H. Wei, J. Li, Z. Wang, X. Tian, A. Pan, and H. Xu, Sci. Rep. **3**, 1967 (2013).
10) S. Kéna-Cohen, P.N. Stavrinou, D.D.C. Bradley, and S.A. Maier, Nano Lett. **13**, 1323 (2013).
11) I. De Leon and P. Berini, Nat. Photonics **4**, 382 (2010).
12) R.F. Oulton, G. Bartal, D.F.P. Pile, and X. Zhang, New J. Phys. **10**, 105018 (2008).
13) V.A. Zenin, S. Choudhury, S. Saha, V.M. Shalaev, A. Boltasseva, and S.I. Bozhevolnyi, Opt. Express **25**, 12295 (2017).
14) P. Berini and I. De Leon, Nat. Photonics **6**, 16 (2011).
15) A. Paul, Y.-R. Zhen, Y. Wang, W.-S. Chang, Y. Xia, P. Nordlander, and S. Link, Nano Lett. **14**, 3628 (2014).
16) M.I. Stockman, Opt. Express **19**, 22029 (2011).
17) M.T. Hill and M.C. Gather, Nat. Photonics **8**, 908 (2014).
18) S.W. Eaton, A. Fu, A.B. Wong, C.-Z. Ning, and P. Yang, Nat. Rev. Mater. **1**, 16028 (2016).
19) K. Leosson, J. Nanophotonics **6**, 061801 (2012).
20) N.K. Dutta and Q. Wang, *Semiconductor Optical Amplifiers* (World scientific, Singapore, 2013) 2nd ed.
21) N.A. Olsson, J. Lightwave Technol. **7**, 1071 (1989).
22) C. Henry, J. Lightwave Technol. **4**, 288 (1986).
23) A.A. Vyshnevyy and D.Y. Fedyanin, Phys. Rev. Appl. **6**, 064024 (2016).
24) I. De Leon and P. Berini, Opt. Express **19**, 20506 (2011).
25) C. Garcia, V. Coello, Z. Han, I.P. Radko, and S.I. Bozhevolnyi, Opt. Express **20**, 7771 (2012).
26) K.C.Y. Huang, M.-K. Seo, T. Sarmiento, Y. Huo, J.S. Harris, and M.L. Brongersma, Nat. Photonics **8**, 244 (2014).
27) D.Y. Fedyanin, A.V. Krasavin, A.V. Arsenin, and A.V. Zayats, Nano Lett. **12**, 2459 (2012).
28) D.A. Svintsov, A.V. Arsenin, and D.Y. Fedyanin, Opt. Express **23**, 19358 (2015).
29) D.Y. Fedyanin and A.V. Arsenin, AIP Conf. Proc. **1291**, 112 (2010).
30) D.Y. Fedyanin, D.I. Yakubovsky, R.V. Kirtaev, and V.S. Volkov, Nano Lett. **16**, 362 (2016).
31) K.M. McPeak, S.V. Jayanti, S.J.P. Kress, S. Meyer, S. Iotti, A. Rossinelli, and D.J. Norris, ACS Photonics **2**, 326 (2015).
32) T.D. Visser, H. Blok, B. Demeulenaere, and D. Lenstra, IEEE J. Quantum Electron. **33**, 1763 (1997).
33) A.A. Vyshnevyy and D.Y. Fedyanin, AIP Conf. Proc. **1874**, 030038 (2017).
34) J.A. Conway, S. Sahni, and T. Szkopek, Opt. Express **15**, 4474 (2007).
35) S. Pasricha and N. Dutt, *On-Chip Communication Architectures: System on Chip Interconnect* (Morgan Kaufmann, Burlington, 2010).
36) A.A. Vyshnevyy and D.Y. Fedyanin, ACS Photonics **3**, 51 (2015).